\documentclass[letter,twocolumn]{jpsj3}
\usepackage{txfonts}
\usepackage{color}

\newcommand{\EF}{E_\mathrm{F}}
\newcommand{\Tc}{T_\mathrm{c}}
\newcommand{\kB}{k_\mathrm{B}}

\newcommand{\centigrade}{$^\circ\mathrm{C}$}
\newcommand{\degree}{$^\circ$}
\newcommand{\ohm}{$\mathrm{\Omega}$}

\newcommand{\STO}{$\mathrm{SrTiO_3}$}

\newcommand{\roottimesroot}[2]{$\sqrt{#1}\times\sqrt{#2}$}

\title{Non-charge-transfer origin of $T_\mathrm{c}$ Enhancement in a Surface Superconductor Si(111)-($\sqrt{7}\times\sqrt{3}$)-In with Adsorbed Organic Molecules}
\author{Kenta Yokota$^{1,2}$, Shunsuke Inagaki$^3$, Wenxuan Qian$^1$, Ryohei Nemoto$^2$, Shunsuke Yoshizawa$^4$, Emi Minamitani$^{5,6}$, Kazuyuki Sakamoto$^{3,7,8}$, and Takashi Uchihashi$^{1,2}$\thanks{UCHIHASHI.Takashi@nims.go.jp}}
\inst{$^1$Graduate School of Science, Hokkaido University, Kita-10 Nishi-8, Kita-ku, Sapporo 060-0810, Japan \\
$^2$International Center for Materials Nanoarchitectonics (WPI-MANA), National Institute for Materials Science, Namiki, Tsukuba, Ibaraki 305-0044, Japan \\
$^3$Department of Applied Physics, Osaka University, Osaka 565-0871, Japan\\
$^4$Research Center for Advanced Measurement and Characterization, National Institute for Materials Science, Sengen, Tsukuba, Ibaraki 305-0047, Japan \\ 
$^5$ Institute for Molecular Science, National Institutes of Natural Sciences, Myoudaiji Campus, 38 Nishigo-Naka, Myodaiji, Okazaki, Aichi 444-8585, Japan\\
$^6$ SANKEN, Osaka University, 8-1 Mihogaoka, Ibaraki, Osaka 567-0047, Japan\\
$^7$ Spintronics Research Network Division, Institute for Open and Transdisciplinary Research Initiatives, Osaka University, Osaka 565-0871, Japan\\
$^8$ Center for Spintronics Research Network, Osaka University, Toyonaka, Osaka 560-8531, Japan.
} 

\abst{The effects of adsorption of Zn-phthalocyanine (ZnPc) molecules on the superconductivity of the Si(111)-(\roottimesroot{7}{3})-In surface are studied through transport measurements under ultrahigh vacuum environment.
The ZnPc molecules are found to increase the transition temperature $\Tc$ by 11\% at maximum, which is about 2.7 times the $\Tc$ increase previously reported using CuPc. 
By contrast, angle-resolved photoemission spectroscopy measurements and \textit{ab initio} calculations show that charge transfer from the In atomic layers to ZnPc is substantially smaller than that to CuPc. 
This clearly shows that charge transfer should be excluded as the origin of the increase in $\Tc$. 
The push-back effect induced by physical adsorption of molecules is discussed as a possible mechanism for the $\Tc$ enhancement.}

\begin{document}
\maketitle
Recently, two-dimensional (2D) superconductors with atomic-scale thickness and high crystallinity have been discovered and attracted much attention \cite{Qin_Pb2ML,Zhang_PbIn1ML,Uchihashi_InR7R3Super,Wang_SingleFeSe,Ye_MoS2Super,Ichinokura_SuperGraphene,Cao_TBGSuper,Yu_BSCCOAL,Uchihashi_2DSuperReview}. 
One of the important characteristics of such atomic-layer superconductors is their sensitivity to surfaces and interfaces. 
For example, in a single-layer FeSe grown on a \STO substrate, the superconducting transition temperature ($\Tc$) rises from the bulk value of 8 K to over 40 K, due to the effect of carrier doping from the substrate or alkali metals \cite{He_FeSeFS,Liu_FeSeMicro4pp,Miyata_FeSeKdope}. 
In addition, superconductivity can be induced at the interfaces of insulating compounds and in twisted bilayer graphene by carrier doping using gate electrodes\cite{Ye_MoS2Super,Saito_EDLSuper2D,Cao_TBGSuper}.

Regarding metal atomic layers epitaxially grown on semiconductor surfaces \cite{Qin_Pb2ML,Zhang_PbIn1ML,Uchihashi_InR7R3Super,Yamada_InR7R3Magneto,Ming_SnSiSuper}, carrier doping with a gate-electrode is not straightforward because the atomic layer structure would easily be destroyed at the interface. 
However, by utilizing the weak van der Waals-like interaction between organic molecules and metal surfaces, an ideal interface can be realized (Fig. 1(a)). 
Recently, adsorption of Cu-phthalocyanine (CuPc) molecule was reported to increase the $\Tc$ of a surface superconductor Si(111)-(\roottimesroot{7}{3})-In (referred to as (\roottimesroot{7}{3})-In) by up to about 4\% \cite{Yoshizawa_PcInSi}. 
In this system, charges corresponding to 1.6 electrons per molecule are transferred from In atoms to CuPc molecules, and the increase in $\Tc$ was ascribed to this ``hole doping''. 
However, similar experiments using $\mathrm{F_{16}CuPc}$ molecule resulted in a decrease in $\Tc$ although a larger hole doping is expected due to its strong acceptor character \cite{Sumi_F16CuPc}. 
As exemplified by this, the mechanism for $\Tc$ modulation in atomic-layer superconductors with adsorbed organic molecule has not been fully clarified yet.

In the present study, we assemble ZnPc molecular layers on Si(111)-(\roottimesroot{7}{3})-In and measure the change in $\Tc$ by transport measurements. 
The adsorption of ZnPc molecules increases $\Tc$ by up to 11\%, which is about 2.7 times the corresponding value reported for CuPc. 
By contrast, the charge transfer from the In atomic layers to ZnPc is found to be significantly smaller than that to CuPc by angle-resolved photoemission spectroscopy (ARPES) measurements and \textit{ab initio} calculations.
This clearly shows that charge transfer is not the origin of the observed increase in $\Tc$ in this system. 
We discuss that the $\Tc$ rise observed here may be attributed to the push-back effect, which is commonly observed for surface adsorption of organic molecule .

The surface superconductor (\roottimesroot{7}{3})-In consists of double atomic layers of bulk In(001) planes epitaxially grown on a Si(111) substrate, where a lattice mismatch leads to formation of a \roottimesroot{7}{3} superstructure \cite{Kraft_R7R3,Rotenberg_R7R3,Park_InSiDL,Uchida_NewInSi,Shirasawa_SXRDInSi}. 
A pristine (\roottimesroot{7}{3})-In undergoes a superconducting transition at $T\approx 3$ K \cite{Zhang_PbIn1ML,Uchihashi_InR7R3Super,Yamada_InR7R3Magneto,Uchihashi_InR7R3Resistive,Yoshizawa_InVortex}. 
In this study, ZnPc was adopted as the adsorbed organic molecule. 
ZnPc has a structure in which a metal ion $\mathrm{Zn^{2+}}$ is coordinated to the center of the phthalocyanine framework \cite{Liao_MPcs}. 
Since it has an electronic configuration similar to that of CuPc where $\mathrm{Cu^{2+}}$ is coordinated to the center (Fig. 1(b)), their comparison should be insightful. 
After flash-cleaning of a Si(111) substrate at 1250\centigrade\ in an ultrahigh vacuum (UHV) environment ($P<1\times 10^{-7}$ Pa), In was deposited and the sample was annealed around 300\centigrade. 
This routinely leads to the formation of a (\roottimesroot{7}{3})-In surface.
 The surface structure and crystallinity of the prepared sample were characterized through low-energy electron diffraction (LEED) and scanning tunneling microscopy (STM) at room temperature. 
 ZnPc (purity $> 98\%$) was deposited on the sample from a crucible by resistive heating. 
 The ZnPc coverage $\theta$ was estimated with a quartz crystal microbalance and was further checked with STM. 
 Here, one monolayer (ML) is defined as the coverage of the molecular layer fully covering the surface.

\begin{figure}[tb]
\begin{center}
\includegraphics[width=7.5cm]{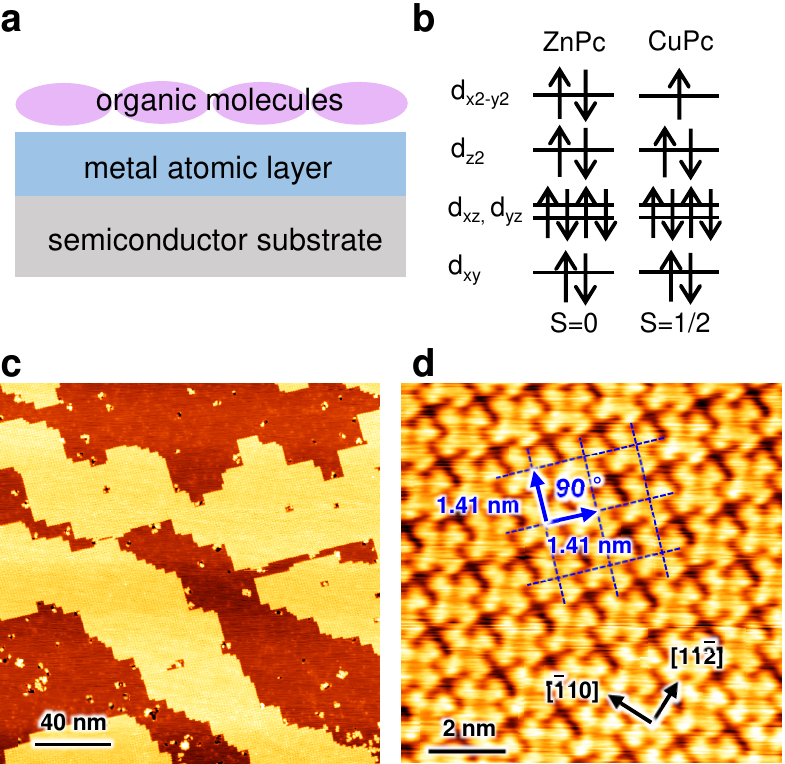}
\caption{(Color online) (a) Schematic diagram of atomic-layer superconductor and organic molecular layer grown on a semiconductor substrate. (b) Electron and spin states of the coordinated metal ions of ZnPc and CuPc in an isolated system. (c) (d) STM images of ZnPc layers grown on a (\roottimesroot{7}{3})-In surface. Coverage: $\theta = 0.5$ ML, sample voltage: $V_s= -2.0$ V, tunnel current: $I_t=10$ pA. (c) Image size: $200\times 200 \ \mathrm{nm^2}$. (d) Image size: $10\times 10 \ \mathrm{nm^2}$. The black arrows indicate the $[11\overline{2}]$ and $[\overline{1}10]$ directions of the Si substrate, while the blue (gray) arrows the unit vector of the ZnPc lattice.}
\label{fig:SchematicSTM}
\end{center}
\end{figure}

Transport measurements were carried out to determine the superconducting transition temperature $\Tc$ of the sample. 
The temperature dependence of the sample resistance was acquired by the four-terminal method using a home-built UHV-compatible cryostat \cite{Uchihashi_InR7R3Super,Uchihashi_InR7R3Resistive,Yoshizawa_Shield,Yoshizawa_DynamicRashba}. 
A non-doped Si wafer (resistivity $\rho>1000$ \ohm cm) was used as the substrate to avoid an leak current at low temperatures. 
After the (\roottimesroot{7}{3})-In surface was prepared, the current path and electrode area were defined with $\mathrm{Ar^{+}}$ sputtering through a shadow mask \cite{Uchihashi_InR7R3Resistive,Yoshizawa_DynamicRashba}, which enables quantitative transport measurements.
The sample was cooled to 1.7 K by liquid helium pumping. 
First, the resistance of the pristine (\roottimesroot{7}{3})-In was measured. 
ZnPc layers with a sub-ML coverage was then grown on the same sample and the measurement was taken again; this process was repeated until the coverage reached 1.5 ML. 
Separately, STM and ARPES measurements were conducted at low temperatures to investigate the molecular assembly structure and the charge transfer due to ZnPc adsorption, respectively. 
A highly doped n-type Si wafer ($\rho < 0.01 \ \mathrm{\Omega cm}$) was used for the former, while a moderately doped n-type Si wafer ($\rho = 1-5 \ \mathrm{\Omega cm}$) was used for the latter.
The amount of the charge transfer and spin magnetic moment in the molecule were also obtained by \textit{ab initio} calculations, and the density of states (DOS) of (\roottimesroot{7}{3})-In near the Fermi level was also calculated \cite{SM1}. 

Figure 1(c) shows the STM image of ZnPc layers ($\theta=0.5$ ML) grown on the (\roottimesroot{7}{3})-In surface. 
The ZnPc layers consist of 1 ML-thick islands with their edges extending along two orthogonal directions. 
The growth of molecules follows the layer-by-layer mode at least up to 2 ML coverage. 
Fig. 1(d) shows a representative STM image of a ZnPc layer taken with the molecular resolution. 
The black arrows indicate the $[11\overline{2}]$ and $[\overline{1}10]$ directions of the Si lattice. 
These directions were determined by measuring the atomic image of (\roottimesroot{7}{3})-In. 
The ZnPc lattice has a unit cell of a $1.4\ \mathrm{nm} \times 1.4\ \mathrm{nm}$ square and its principal axes are rotated by 45\degree\ with respect to the $[11\overline{2}]$ and $[\overline{1}10]$ directions. 
This structure is identical to that of CuPc layers on the (\roottimesroot{7}{3})-In surface reported previously \cite{Yoshizawa_PcInSi}. 
This observation suggests that ZnPc forms an epitaxial molecular film due to a lattice matching to (\roottimesroot{7}{3})-In as in the case of CuPc. 

Transport measurements were carried out to investigate the effect of ZnPc coverage on the $\Tc$ of (\roottimesroot{7}{3})-In. 
Figure 2(a) plots the temperature dependence of the sheet resistance $R_\mathrm{sheet}$ (\textit{i.e.,} 2D resistivity) observed for different coverages of ZnPc. 
They all exhibit a steep drop at low temperatures due to the superconducting transition.
The figure also shows the results of fitting based on thermal fluctuation theories of 2D superconductivity \cite{Uchihashi_InR7R3Resistive,SM2}.
The fitting range was taken to be $0.6 \ R_\mathrm{n} <R_\mathrm{sheet}< R_\mathrm{n}$, where $R_\mathrm{n}$ is defined as $R_\mathrm{sheet}$ measured at 4.5 K. 
For the pristine sample ($\theta=0.0$ ML), $\Tc=2.81$ K was obtained (solid circles). 
$\Tc$ was found to increase with increasing coverage of ZnPc, reaching the maximal value of $\Tc = 3.13$ K at a coverage of $\theta=1.0$ ML (open squares). 
$\Tc$ showed a decreasing trend for $\theta>1.0$ ML, resulting in $\Tc=3.10$ K at $\theta=1.5$ ML (solid triangles).
We note that, while the resistive transition is very sharp for $\theta=0.0$ ML, its width becomes broadened for $\theta>0.6$ ML.
Accordingly,  the experimental data exhibit noticeable deviations from the fitting curves for $R_\mathrm{sheet}< 0.6 \ R_\mathrm{n}$.
This suggests that the adsorption of ZnPc molecules induces a spatial inhomogeneity of $\Tc$ and that the $\Tc$ determined here is close to the maximal value.

\begin{figure}[tb]
\begin{center}
\includegraphics[width=8.6cm]{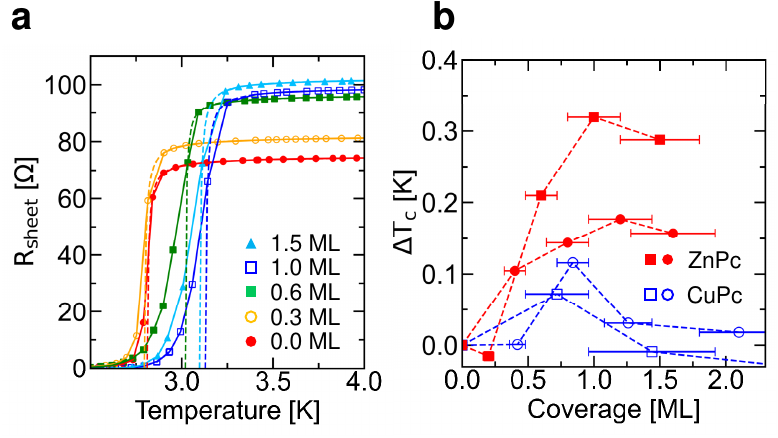}
\caption{(a) (Color online) Temperature dependences of the sheet resistance of (\roottimesroot{7}{3})-In acquired for different coverages of ZnPc. The dashed lines are the fits to thermal fluctuation theories of 2D superconductivity \cite{Uchihashi_InR7R3Resistive,SM2}. (b) Variation of $\Delta \Tc$ as a function of ZnPc coverage (solid circles/squares). The result of the previous experiment using CuPc \cite{Yoshizawa_PcInSi} is also plotted (open circles/squares).}
\label{fig:Transport}
\end{center}
\end{figure}

Figure 2(b) displays $\Delta \Tc$ defined as a variation in $\Tc$ ($\Delta \Tc \equiv \Tc(\theta) - \Tc(\theta=0))$ acquired for two different ZnPc/(\roottimesroot{7}{3})-In samples (solid circles/squares).
 The two plots follow the qualitatively same behavior. 
$\Delta \Tc$ amounts to $0.32$ K at maximum, corresponding to about 11\% of $\Tc$ of (\roottimesroot{7}{3})-In. 
 Figure 2(b) also plots $\Delta \Tc$ induced by CuPc adsorption (open circles/squares) \cite{Yoshizawa_PcInSi}. 
 $\Tc$ also peaks around $\theta=1.0$ ML but only reaches $0.12$ K at maximum. 
Thus, the maximum value of $\Delta \Tc$ for ZnPc is about 2.7 times larger than that for CuPc. 
In the previous work\cite{Yoshizawa_PcInSi}, the increase in $\Tc$ was attributed to charge transfer from the In to the molecular layer (\textit{i.e.}, hole doping). 
If this interpretation is correct, ZnPc should cause a larger charge transfer than CuPc.

To estimate the amount of charge transfer from ZnPc molecule, the change of the Fermi surface of (\roottimesroot{7}{3})-In was studied through ARPES measurements. 
Figure 3(a) shows the result for a pristine (\roottimesroot{7}{3})-In surface.
The color (brightness) of the figure indicates photoelectron intensity, and the $k_x, k_y$ axes correspond to the $[\overline{1}10]$ and $[11\overline{2}]$ directions, respectively. 
The Fermi surface consists of two arc-like structures enclosed by the dahsed lines and butterfly-like structures in between, which agree well with the previous reports \cite{Rotenberg_R7R3,Yoshizawa_PcInSi,Sagehashi_ARPESCuPc,Kobayashi_OAM,Yoshizawa_DynamicRashba}. 
Figure 3(b) shows the result for the (\roottimesroot{7}{3})-In surface covered with $1.5 (\pm 0.5)$ ML-thick ZnPc layer.
The Fermi surface of (\roottimesroot{7}{3})-In is still clearly visible, while the background of photoelectron intensity increases. 
This strongly suggests that the atomic structure of (\roottimesroot{7}{3})-In is preserved under the ZnPc layer and that there exist no strong chemical bonds between ZnPc and In atoms. 
Notably, as shown in Figs. 3(c) and 3(d), the peak positions of momentum distribution curves (MDCs)  are shifted only slightly. 
These arc-like structures are the consequence of back-folding of the circular 2D Fermi surface, which originates from the In conduction electrons \cite{Rotenberg_R7R3}. 
If holes are doped into (\roottimesroot{7}{3})-In by ZnPc adsorption as expected, the radius of the original Fermi surface shrinks. 
This in turn leads to a decrease in the distance between the two arc-like structures \cite{Yoshizawa_PcInSi,Sagehashi_ARPESCuPc}. 
Our result described above contradicts this scenario.

\begin{figure}[tb]
\begin{center}
\includegraphics[width=8.6cm]{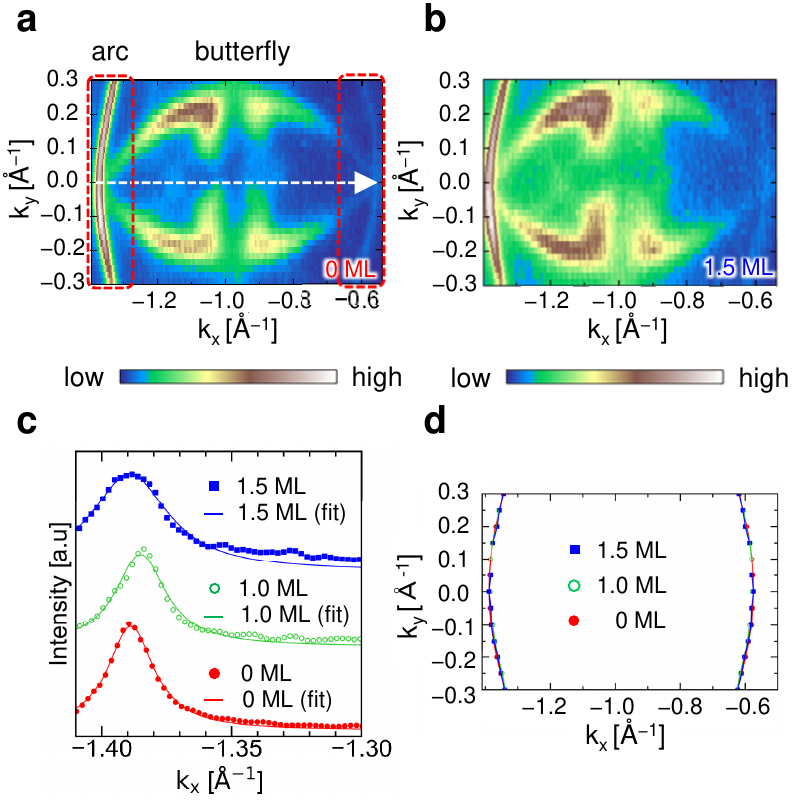}
\caption{(Color online) Photoelectron mapping of the (\roottimesroot{7}{3})-In Fermi surface in momentum space before (a) and after (b) deposition of $ 1.5 (\pm 0.5)$ ML ZnPc. The region encircled by the dashed lines indicates an arc-like Fermi surface. (c)(d) MDCs obtained along the $k_y =0$ line (c) and peak positions of the arc-like Fermi surface of (\roottimesroot{7}{3})-In (d) acquired for different ZnPc coverages. Solid circles: $0$ ML, open circles: $1.0\pm0.3$ ML, solid squares: $1.5\pm0.5$ ML. }
\label{fig:ARPES}
\end{center}
\end{figure}

The amount of charge transfer due to the molecule adsorption is evaluated as follows. 
The maximal peak shifts $-0.0037\pm 0.007 \ \mathrm{\AA }^{-1}$ is detected for $1.0$ ML coverages of ZnPc. 
This leads to a charge transfer per molecule $\Delta n_\mathrm{tr}= -0.31 \pm 0.59 \ e$, where $e$ is the elementary charge (here, the negative sign of $\Delta n_\mathrm{tr}$ corresponds to electron transfer from the In layers to the molecules.) 
Meanwhile, the analogous ARPES measurements revealed that the radius of the (\roottimesroot{7}{3})-In Fermi surface shrank by $0.018 \pm 0.004 \ \mathrm{\AA }^{-1}$ when 1.0 ML-thick CuPc was adsorbed instead of ZnPc \cite{Sagehashi_ARPESCuPc}. 
This leads to $\Delta n_\mathrm{tr} = -1.51 \pm 0.32 \ e$ as a charge transfer due to CuPc adsorption \cite{SM3}.
We also carried out \textit{ab initio} calculations to quantify the charge transfer due to ZnPc.
The result is $\Delta n_\mathrm{tr} = -0.76 \ e$, which is only 47\% of the value $\Delta n_\mathrm{tr} = -1.61 \ e$ previously obtained for CuPc with the same method \cite{Yoshizawa_PcInSi}. 
Table~\ref{tab:TcCT} summarizes the maximal values of $\Delta \Tc$ obtained by transport measurements and the amount of charge transfer deduced from the ARPES measurements and the \textit{ab initio} calculations. 
Both ARPES and \textit{ab initio} calculations show that ZnPc induces less charge transfer than CuPc, which is opposite to the expected trend based on the variations in $\Tc$ \cite{footnote1}. 
Therefore, we conclude that the increase in $\Tc$ is not attributable to charge transfer between the molecules and the In atoms.

\begin{table}[bt]
\caption{List of the maximal changes in $\Tc$ ($\Delta \Tc$) and charge transfers ($\Delta n_\mathrm{tr}$) induced by ZnPc and CuPc molecules. The negative sign of $\Delta n_\mathrm{tr}$ corresponds to electron transfer from the In layers to the molecule.) }
\label{tab:TcCT}
\begin{center}
\begin{tabular}{llcc}
\hline
physical quantity & method & ZnPc & CuPc\\
\hline
$\Delta \Tc$ (K) & transport & 0$.32$ & $0.12$\\
$\Delta n_\mathrm{tr} (e)$ & ARPES & $-0.31\pm 0.59$ & $-1.51\pm 0.32$\\
$\Delta n_\mathrm{tr} (e)$ & \textit{ab initio} calc. & $-0.76$ & $-1.61$\\
\hline
\end{tabular}
\end{center}
\end{table}

At present, the mechanism of the $\Tc$ enhancement is not clear. We first note that the emergence of unconventional superconductivity, which is caused by a strong electron correlation as in high-$\Tc$ cuprates, is very unlikely since the system can be described by a one-electron band picture \cite{Yoshizawa_DynamicRashba,Kobayashi_OAM}. This idea is supported by the observation of the \textit{s}-wave like superconducting energy gap reported previously \cite{Yoshizawa_InVortex}. 
Here we propose the push-back effect as a possible origin based on the standard Bardeen-Cooper-Schrieffer (BCS) theory.
Since the potential barrier for electrons at the metal surface is finite in height, the wavefunctions of conduction electrons penetrate into the vacuum region.
However, when organic molecules are physically adsorbed on the surface, the conduction electrons are pushed back toward the metal nuclei by Pauli repulsion because of overlapping with the molecular wavefunctions. 
This is called the push-back effect (or the Pauli repulsion effect) \cite{Bagus_PauliRepul,Witte_CushionEffect,Toyoda_Pushback}. 
In particular, phthalocyanine molecules adsorbed in the planar configuration are expected to cause a strong effect, because their $\pi$-conjugated orbitals extends perpendicular to the surface. 
This should enhance both the electron DOS per unit volume and the attractive interactions between electrons through an increase in electron-phonon interactions.
According to the BCS theory, $\Tc$ is expressed by the following formula \cite{Tinkham_Textbook}
\begin{equation}
\kB \Tc = 1.13\hbar \omega_c \exp\left( -\frac{1}{N(\EF)V} \right)
\end{equation}
where $\omega_c$ is the Debye phonon frequency, $N(\EF)$ is the DOS per unit volume at the Fermi level $\EF$, and $V$ is the attractive potential between electrons. 
Therefore, $\Tc$ is expected to rise when $N(\EF)V$ increases due to the push-back effect. 
Using the typical value $N(\EF)V \sim 0.2$ for BCS superconductors, an increase in $N(\EF)V$ by 2\% enhances $\Tc$ by 10\%. 
Thus, this mechanism is considered realistic.
We note that $\Tc$ variations based on an analogous mechanism has been predicted for the superconductivity of monolayer graphite intercalation compounds \cite{Profeta_LiGML}.

\begin{figure}[tb]
\begin{center}
\includegraphics[width=8.6cm]{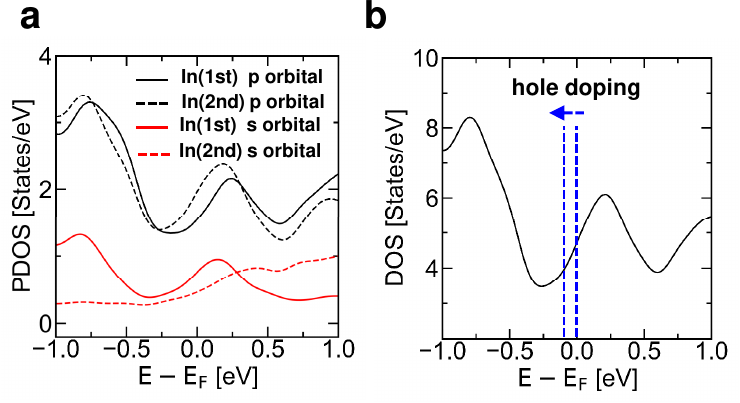}
\caption{(Color online) Energy dependence of DOS of (\roottimesroot{7}{3})-In obtained by \textit{ab initio} calculations. (a) DOS projected onto individual orbitals of In atoms. Black solid/dashed lines: $p$-orbitals of the first (top) and second (bottom) In layers. Red (gray) solid/dashed lines: $s$-orbitals of the first and second In layers.  (b) Summation of PDOS in (a).}
\label{fig:DOS}
\end{center}
\end{figure}

Since the push-back effect is a universal phenomenon caused by physically adsorbed organic molecules, it is expected to appear commonly for ZnPc and CuPc. 
Then, what is the cause of the difference in $\Delta\Tc$ induced by these molecules? 
According to Park \textit{et al.}\cite{Park_InSiDL,Park_InSiHex}, the DOS of (\roottimesroot{7}{3})-In is an increasing function of the energy $E$ near the Fermi level $\EF$. 
Therefore, a hole doping should lead to a decrease in $N(E)$ and consequently to an decrease in $\Tc$. 
Since the hole doping effect is larger for CuPc than for ZnPc, the $\Tc$ decrease is more significant for CuPc. 
To further check this idea, we independently performed \textit{ab initio} calculations for (\roottimesroot{7}{3})-In to obtain the projected DOS (PDOS) for individual atomic orbitals.
Figures 4(a) and 4(b) show the calculated PDOS, where the former refers to the DOS projected onto the $s, p$ orbitals of In atoms in the first (top) and second (bottom) layers and the latter to the summation of individual contributions. 
Since all of them are increasing functions of $E$ near $\EF$, the above argument should to be applicable irrespective of the degree of contribution of each orbital to superconductivity.

Finally, we consider the possibility that localized spins in ZnPc and CuPc, if exist, may suppress superconductivity  \cite{Yoshizawa_PcInSi,Uchihashi_FePcInSi}. 
As depicted in Fig. 1(b), ZnPc has no total spin in the isolated state.
Our \textit{ab initio} calculations confirmed that spin magnetic moment $m_\mathrm{s}$ is equal to $0.00 \ \mu_B$ ($\mu_B$: Bohr magneton) even after adsorption on the (\roottimesroot{7}{3})-In surface.
By contrast, CuPc has a total spin $\hbar/2$ in the isolated state and retains $m_\mathrm{s}=0.29 \ \mu_B $ after adsorption on (\roottimesroot{7}{3})-In.
Nevertheless, the magnetic moment induced in the In atomic layer is $m=0.00 \ \mu_B $ \cite{Yoshizawa_PcInSi}. 
This means that there is no exchange interaction between the spins of CuPc and those of conduction electrons in the In layer. 
This is because the $d_{x^2-y^2}$ orbital accommodating the spins of CuPc hardly overlaps with the wavefunctions of the In atoms due to its planar form.
Therefore, the effect of localized spins on superconductivity can be neglected for both ZnPc and CuPc.

In summary, we have clarified that the charge transfer between organic molecules and In atoms should be excluded as the origin of the $\Tc$ enhancement in (\roottimesroot{7}{3})-In. 
On the contrary, the charge transfer is likely to lower $\Tc$. 
While the origin of the $\Tc$ increase in this surface system is not clear at present, we have proposed the push-back effect due to the organic molecules as a plausible mechanism.
The present study may provide an important clue for understanding the $\Tc$ modulations, which have been observed in many kinds of surface/interface 2D superconductors.

\begin{acknowledgment}
The financial support from the following is acknowledged: JSPS Kakenhi Grant No. 22H01961, 20H05621, 20H02707, and World Premier International Research Center (WPI) Initiative on Materials Nanoarchitectonics (T. U.), JSPS Kakenhi Grant No. 22H01957, 19H02592, and the Spintronics Research Network of Japan (Spin-RNJ) (K. S).
\end{acknowledgment}

\bibliographystyle{/texlive/texmf-local/tex/jpsj3/jpsj} 

\bibliography{MyEndNoteLibrary}

\begin{thebibliography}{10}
\bibitem{footnote1}
The samples used in the transport and ARPES experiments should have the same carrier densities in the In layers despite the different dopant levels in the Si substrates. This is supported by the fact that the radius of the Fermi surface of (\roottimesroot{7}{3})-In obtained with a high-doped Si wafer ($\rho < 0.01 \ \mathrm{\Omega cm}$)\cite{Kobayashi_OAM} is equal to that obtained in the present study ($\rho =1-5 \ \mathrm{\Omega cm}$) within the experimental error.
\end{thebibliography}

\end{document}